\documentclass[aps,twocolumn,showpacs,superscriptaddress]{revtex4-2}

\usepackage[all, warning]{onlyamsmath}
\usepackage{amssymb, amsmath}
\usepackage{graphicx}
\usepackage{dcolumn}
\usepackage{bm}
\usepackage[figuresright]{rotating}
\usepackage{fancyhdr}
\usepackage{enumitem}
\usepackage{siunitx}
\usepackage{indentfirst}
\usepackage{xspace}
\usepackage{epsfig}
\usepackage{grffile}
\usepackage[T1]{fontenc}
\usepackage[utf8]{inputenc}
\usepackage[export]{adjustbox}
\usepackage{dsfont}

\DeclareMathOperator*{\argmax}{\arg\max} 

\begin{document}

\title{Machine-learning approach for operating electron beam at KEK $e^-$/$e^+$ injector Linac}

\author{Gaku~Mitsuka}
\email[]{gaku.mitsuka@kek.jp}
\affiliation{KEK, Oho, Tsukuba, Ibaraki 305-0801, Japan}
\affiliation{SOKENDAI, Shonan Village, Hayama, Kanagawa 240-0193, Japan}

\author{Shinnosuke~Kato}
\affiliation{The University of Tokyo, Bunkyo, Tokyo 113-0033, Japan}

\author{Naoko~Iida}
\affiliation{KEK, Oho, Tsukuba, Ibaraki 305-0801, Japan}
\affiliation{SOKENDAI, Shonan Village, Hayama, Kanagawa 240-0193, Japan}

\author{Takuya~Natsui}
\affiliation{KEK, Oho, Tsukuba, Ibaraki 305-0801, Japan}
\affiliation{SOKENDAI, Shonan Village, Hayama, Kanagawa 240-0193, Japan}

\author{Masanori~Satoh}
\affiliation{KEK, Oho, Tsukuba, Ibaraki 305-0801, Japan}
\affiliation{SOKENDAI, Shonan Village, Hayama, Kanagawa 240-0193, Japan}

\date{\today}

\begin{abstract}
In current accelerators, numerous parameters and monitored values are to be
adjusted and evaluated, respectively. In addition, fine adjustments are required
to achieve the target performance. Therefore, the conventional
accelerator-operation method, in which experts manually adjust the parameters,
is reaching its limits. We are currently investigating the use of machine
learning for accelerator tuning as an alternative to expert-based tuning. In
recent years, machine-learning algorithms have progressed significantly in terms
of speed, sensitivity, and application range. In addition, various libraries are
available from different vendors and are relatively easy to use. Herein, we
report the results of electron-beam tuning experiments using Bayesian
optimization, a tree-structured Parzen estimator, and a covariance
matrix-adaptation evolution strategy. Beam-tuning experiments are performed at
the KEK $e^-$/$e^+$ injector Linac to maximize the electron-beam charge and
reduce the energy-dispersion function. In each case, the performance achieved is
comparable to that of a skilled expert.
\end{abstract}

\maketitle

\section{Introduction}\label{sec:introduction}
To improve or maintain the high performance of modern accelerators, dozens or
even hundreds of parameters must be optimized to accommodate the volatile
conditions. Values monitored to determine the success or failure of the
optimization include those of the beam orbit, beam charge, energy-dispersion
function, emittance, and charge loss in each accelerator sector. Hundreds of
values are monitored. Determining the operating parameters, such as the magnetic
field, based solely on the beam dynamics is typically challenging. For example,
the KEK $e^-$/$e^+$ injector Linac (referred to as the KEK Linac) has no
monitors to diagnose the beam energy in each sector, and the energy gains of
individual RF cavities are accurately determined only occasionally.
In addition, changes in the environmental temperature affect the RF system and
set energy drifts.
Therefore, the actual operation requires beam-parameter optimization while the
beam conditions are monitored.
Hitherto, operation experts have optimized the beam parameters based on their
knowledge and experience. Complex and sensitive accelerator operations, such
live optimizations based on expert inputs, may be time consuming reproduce, even
if the results satisfy the required criteria.
In particular, in the case of Linac accelerating a beam in a single pass, a
self-feedback mechanism does not exist, unlike the ring where the beam orbits.
Thus, the beam condition cannot be reproduced easily even if the same operating
parameters are set.

Accelerator tuning using machine learning has recently garnered attention as an
alternative to expert-dependent optimization. Machine learning has progressed
significantly in terms of speed, sensitivity, and application range since the
2010s. Various libraries are available from different vendors and are relatively
easy to use.
For example, a neural network approach learns the response between the applied
current of a coil (input) and a beam orbit (output) over weeks to months.
Subsequently, it predicts the best applied current to achieve an optimal beam
orbit.
Because neural networks are based on supervised learning, their regression
calculation accuracy is generally higher than that of unsupervised learning.
Additionally, they can detect (classify) anomalies. However, the overtraining
problem, in which the regression accuracy decreases significantly when the test
situation differs from the training data, must be
addressed~\cite{bishop2006pattern}. In certain accelerator applications,
differences in the environmental temperature and other time-dependent drifts may
deteriorate the regression performance.

Deep learning~\cite{Goodfellow-et-al-2016}, which a new machine-learning method
that uses autoencoders and fine tuning, may offer better predictive ability and
solve overlearning problems. Nevertheless, Mishra \textit{et
al.}~\cite{PhysRevAccelBeams.24.114601} highlighted that even predictions from a
trained neural network may contain significant errors and uncertainties owing to
bias, noise, and other inevitable factors. The inadequate interpretability of
deep-learning models may worsen the effect of such errors and uncertainties on
particle-accelerator applications. To the best of our knowledge, deep-learning
applications for accelerator tuning have not been reported.

The objective of this study is to realize accelerator tuning using unsupervised
parameter-optimization algorithms. Bayesian
optimization~\cite{garnett_bayesoptbook_2023} is a parameter-optimization
algorithm based on the Gaussian process~\cite{10.7551/mitpress/3206.001.0001}
and Bayesian decision theory~\cite{Berger:1327974}. It performs unsupervised
learning through multiple changing parameters and obtains response trials.
Because it does not use supervised data, it is less susceptible to environmental
changes, such as temperature drifts. Nevertheless, multiple trials are required
to improve the prediction accuracy. In accelerator tuning, in which the cost of
a single trial cannot be disregarded, one should avoid performing numerous
trials.
Owing to these conflicting demands, we evaluated Bayesian optimization for
actual beam tuning to determine whether the target perperformance was achieved,
the number of trials required for the specified parameters, and whether multiple
goals (e.g., beam charge vs. dispersion function) were achieved simultaneously.
In addition to Bayesian optimization, we evaluated other parameter-optimization
algorithms, i.e., the tree-structured Parzen estimator
(TPE)~\cite{10.5555/2986459.2986743} and the covariance matrix-adaptation
evolution strategy (CMA-ES)~\cite{hansen2023cma}, by comparing their
characteristics and tuning results.
We performed a beam-tuning experiment at the KEK Linac in June 2023 using a
parameter-optimization library name
\textsc{optuna}~\cite{10.1145/3292500.3330701}. \textsc{optuna} internally
implements not only the BoTorch~\cite{balandat2020botorch} (Bayesian
optimization), TPE, and CMA-ES algorithms, but also other single- and
multiobjective optimization algorithms.
The results of beam tuning using various optimization algorithms provide
essential guidelines for future machine-learning applications associated with
particle accelerators.

The remainder of this paper is organized as follows: Section~\ref{sec:single}
describes the flow of single-objective optimization using Bayesian optimization,
the TPE, and the CMA-ES. Section~\ref{sec:single-test} reports the experimental
results of the single-objective optimization performed at the KEK Linac.
Section~\ref{sec:multi} describes the flow of multiobjective optimization.
Section~\ref{sec:multi-test} details an actual beam experiment performed at the
KEK Linac. Finally, Section~\ref{sec:conclusions} concludes the paper.

%
%
\section{Single-objective optimization}\label{sec:single}

For simplicity, we define \textit{parameter optimization} as the process of
identifying optimal \textit{parameter} values (e.g., applied currents of
steering magnetic coils) to \textit{minimize} the \textit{objective function}
(e.g., dispersion function). If the parameters are to be optimized by
\textit{maximizing} the objective function (e.g., the beam charge), then the
discussion in this section can be applied by reversing the sign of the evaluated
value.

\subsection{Bayesian optimization}\label{sec:botorch}

Bayesian optimization is an iterative strategy for the global optimization of
black-box functions using a stochastic process known as the Gaussian process. A
detailed description of the Gaussian processes is provided in
Ref.~\cite{10.7551/mitpress/3206.001.0001}. In this study, we used the BoTorch
algorithm~\cite{balandat2020botorch} via
\textsc{optuna}~\cite{10.1145/3292500.3330701} for Bayesian optimization.

In Bayesian optimization, the objective function $f(x)$ for each putative input
location $x$ is assumed to adhere to a Gaussian process $\mathcal{GP}(0,k)$ with
a mean function of $0$ and a covariance function $k$. The covariance function is
similarly known as the kernel function, and the Matern kernel function is used
in this study.
\begin{align}
k_\nu(x, x') &= \frac{2^{1-\nu}}{\Gamma(\nu)}(\sqrt{2\nu}d)^\nu
K_\nu(\sqrt{2\nu}d) \\ \nonumber
d &= (x-x')^T\Theta^2(x-x').
\end{align}
Here, $K_\nu$ is the modified Bessel function of the second kind. The parameter
$\nu$ determines the smoothness of the function, and we use $\nu=5/2$ in this
study. The lengthscale parameter $\Theta$ scales the kernel function and is
estimated from data.
The automatic relevance determination technique is employed in this study to set
the lengthscale parameter $\Theta$ very long, resulting in a covariance function
effectively removing irrelevant dimensions~\cite{garnett_bayesoptbook_2023,
10.7551/mitpress/3206.001.0001}.

In a Gaussian process, the observed value can be written as
$y=f(x)+\varepsilon$. A noise $\varepsilon$ is assumed adhere to $\varepsilon
\sim \mathcal{N}(0, \sigma^2)$, where a constant variance $\sigma^2$ is
independent of the location $x$ and is inferred in the algorithm. Under these
assumptions, we obtain the following probability model between location $x$ and
observed value $y$:
\begin{equation}
p(y \mid x,\mathcal{H}_t) = \mathcal{N}(\mu_t(x,\mathcal{H}_t), \sigma_t(x,\mathcal{H}_t)^2).
\label{eq:prob_pyx}
\end{equation}
$\mathcal{H}_t = \{(x_i,y_i)\}_{i=1}^{t-1}$ is the history of pairs of location
$x$ and observed value $y$ up to the $t-1$-th trial. For the specific forms of
$\mu_t$ and $\sigma_t$, please refer to
Ref.~\cite{10.7551/mitpress/3206.001.0001}.

The acquisition function in Bayesian optimization is defined as a real-valued
function $\alpha(x, \mathcal{H}_t)$ in the space of objective function
$\mathcal{X}$. In each trial, the location $x$ that maximizes the acquisition
function $\alpha$ is selected, which is then input to the objective function
$f(x)$ to obtain the observed value $y$.
The expected improvement (EI) is applied in the acquisition function used in
this study.
\begin{align}
x_t &\in \argmax_{x' \in \mathcal{X}} \alpha(x'; \mathcal{H}_t) \\ \nonumber
\alpha_\textrm{EI}(x\,; \mathcal{H}_t) &= \int \max(y^\ast_t - y,0)p(y \mid x, \mathcal{H}_t) dy.
\end{align}
Specifically, the EI represents the expected value for the conditional
probability distribution $p(y \mid x, \mathcal{H}_t)$ and the extent by which
the observed value $y$ improves from the currently obtained minimum $y$ value
(denoted $y^\ast_t$) based on a location $x$ selected; and $y^\ast_t$ is a
constant determined based on the history $\mathcal{H}_t$.

\subsection{TPE}\label{sec:tpe}

Unlike Bayesian optimization with Gaussian processes, which directly models $p(y
\mid x,\mathcal{H}_t)$, as shown in Eq.~(\ref{eq:prob_pyx}), the
TPE~\cite{10.5555/2986459.2986743} models $p(x \mid y, \mathcal{H}_t)$ and
$p(y)$ and uses them to calculate the EI. The conditional probability
distribution can be modeled using two densities, as follows:
\begin{equation}
p(x \mid y, \mathcal{H}_t) =
\begin{cases}
    l(x) & (y < y^\ast_t) \\
    g(x) & (y \geq y^\ast_t).
\end{cases}
\label{eq:parzen}
\end{equation}
The distribution exhibited by $x$ when the observed value $y$ is lower than
$y^\ast_t$ is denoted as $l(x)$, and the distribution when $y$ is larger than
$y^\ast_t$ as $g(x)$. The two distributions in Eq.~(\ref{eq:parzen}) are
obtained using kernel density estimators~\cite{watanabe2023treestructured}.
$y^\ast_t$ shall be provided to satisfy the probability $\gamma = p(y<y^\ast_t
\mid \mathcal{H}_t)$ for the predefined threshold $\gamma$ ($0<\gamma<1$).

Using Eq.~(\ref{eq:parzen}), we can obtain the EI from the TPE.
\begin{align}
\label{eq:tpeei1}
\alpha_\textrm{EI}(x\,; \mathcal{H}_t) &= \int_{-\infty}^{\infty} \max(y^\ast_t-y,0)p(y \mid x, \mathcal{H}_t) dy \\ \nonumber
&= \int_{-\infty}^{y^\ast_t} (y^\ast_t-y)\frac{p(x \mid y, \mathcal{H}_t)p(y \mid \mathcal{H}_t)}{p(x \mid \mathcal{H}_t)} dy \\ \nonumber
&= \frac{l(x)}{p(x \mid \mathcal{H}_t)} \int_{-\infty}^{y^\ast_t} (y^\ast_t-y) p(y \mid \mathcal{H}_t) dy,
\end{align}
where the normalization in the denominator is written as
\begin{align}
\label{eq:tpeei2}
p(x \mid \mathcal{H}_t) &= \int p(x \mid y, \mathcal{H}_t)p(y \mid \mathcal{H}_t) \\ \nonumber
&= \gamma l(x) + (1-\gamma)g(x).
\end{align}
The final integral in Eq.~(\ref{eq:tpeei1}) is independent of $x$. Thus, the
specific form of $p(y \mid \mathcal{H}_t)$ need not be considered when
maximizing $\alpha_\textrm{EI}(x\,; \mathcal{H}_t)$.
Finally, we approximate Eq.~(\ref{eq:tpeei1}) as
\begin{align}
\label{eq:tpeei3}
\alpha_\textrm{EI}(x\,; \mathcal{H}_t) &= \frac{l(x)}{p(x \mid \mathcal{H}_t)} \int_{-\infty}^{y^\ast_t} (y^\ast_t-y) p(y \mid \mathcal{H}_t) dy \\ \nonumber
&\propto \frac{l(x)}{\gamma l(x) + (1-\gamma)g(x)} \\ \nonumber
&= \left( \gamma + \frac{g(x)}{l(x)}(1-\gamma) \right)^{-1}.
\end{align}
Eq.~(\ref{eq:tpeei3}) indicates that the $x$ value maximizing
$\alpha_\textrm{EI}(x\,; \mathcal{H}_t)$ is the location that minimizes the
density ratio $g(x)/l(x)$.

\subsection{CMA-ES}\label{sec:cmaes}

The CMA-ES~\cite{hansen2023cma} is an evolutionary computational algorithm for
continuous optimization problems. In each iteration (\textit{generation},
denoted as $g$), new candidate solutions (\textit{individuals}, denoted as $x$)
are generated following the multivariate normal distribution determined by the
parental individuals. The set of individuals is known as the
\textit{population}, and the \textit{population size} in each generation is
denoted by $\lambda$.
The algorithm comprises the following steps:
\begin{enumerate}
  \item Generate $\lambda$ candidate solutions (individuals) based on the
  multivariate normal distribution $\mathcal{N}(m, \sigma^2 C)$ and calculate
  the objective function for each individual.

  \item Among the $\lambda$ individuals generated, extract $\mu (\mu<\lambda)$
  individuals with the highest-ranking objective functions and update the mean
  vector $m$ by multiplying the $\mu$ individuals by their weights.

  \item Update the variance parameters (i.e., $\sigma$ and $C$) of the
  multivariate normal distribution based on the isotropic and anisotropic
  evolution paths.
  
  \item Repeat steps 1--3.
\end{enumerate}

\textit{Step 1:}
The $k$-th individual $x_k^{(g+1)}$ in generation $g+1$ can be determined based
on the multivariate normal distribution $x_k^{(g+1)} \sim \mathcal{N}(m^{(g)},
(\sigma^{(g)})^2 C^{(g)})$. The mean vector $m^{(g)}$, step size $\sigma^{(g)}$,
and covariance matrix $C^{(g)}$ are parameters constructed from the generation
$g$. The objective function for $x_k^{(g+1)}$ is denoted as $f(x_k^{(g+1)})$.

\textit{Step 2:}
Define $x_{i:\lambda}^{(g)}$ as the $\lambda$ individuals in generation $g$
sorted in the ascending order of the corresponding objective functions. The mean
vector $m^{(g)}$ is written as
\begin{equation}
m^{(g)} = \sum_{i=1}^\mu w_i x_{i:\lambda}^{(g)},
\end{equation}
where $w_i$ is the weight of each individual and satisfies the relation
\begin{equation}
\sum_{i=1}^\mu w_i = 1 \hspace{5mm}(w_1 \geq w_2 \geq \cdots \geq w_\mu \geq 0).
\end{equation}
The specific expression for $w_i$ is provided in Appendix A of
Ref.~\cite{hansen2023cma}. In this study, $\mu = \lambda/2$ is used.

\textit{Step 3:}
The step size $\sigma^{(g)}$ used to search for the ($g+1$)th generation is
obtained by updating the previous step size $\sigma^{(g-1)}$.
\begin{equation}
\sigma^{(g)} = \sigma^{(g-1)} \exp\left[ \frac{c_\sigma}{d_\sigma} \left( \frac{\lVert p_\sigma^{(g)} \rVert}{E\lVert \mathcal{N}(0,I) \rVert} \right) -1 \right].
\end{equation}
For the specific forms of $c_\sigma$, $d_\sigma$, $p_\sigma^{(g)}$, and $E\lVert
\mathcal{N}(0,I) \rVert$, see Appendix A in Ref.~\cite{hansen2023cma}. Next, we
construct a covariance matrix $C$ that characterizes the population distribution
in each generation. The $C^{(g)}$ used for generation $g+1$ is obtained by
updating $C^{(g-1)}$.
Because the input space for new individuals can be changed adaptively through
$\sigma^{(g)}$ only, near-optimal solutions are obtained even without $C^{(g)}$.
However, an isotropic search without $C^{(g)}$ is inefficient because the sphere
space increases exponentially in proportion to the number of input dimensions.
Thus, the isotropic search becomes inefficient, particularly when the
sensitivity to the objective function differs significantly among the input
parameters. The covariance matrix is updated such that the input space expands
in the direction in which the sensitivity of the objective function increases.
A detailed discussion regarding updates to the covariance matrix is provided in
Sec.~3 of Ref.~\cite{hansen2023cma}.

The covariance matrix $C^{(0)}$ is initialized with a unit matrix $I$. The user
should set the normal distribution center $m^{(0)}$ and step size $\sigma^{(0)}$
based on the problem to be addressed.

%
%
\section{Application to beam-charge maximization}\label{sec:single-test}

\subsection{A. Experimental setup at KEK Linac}\label{sec:setup1}

\begin{figure*}[hbt]
 \centering
 \includegraphics[width=\linewidth]{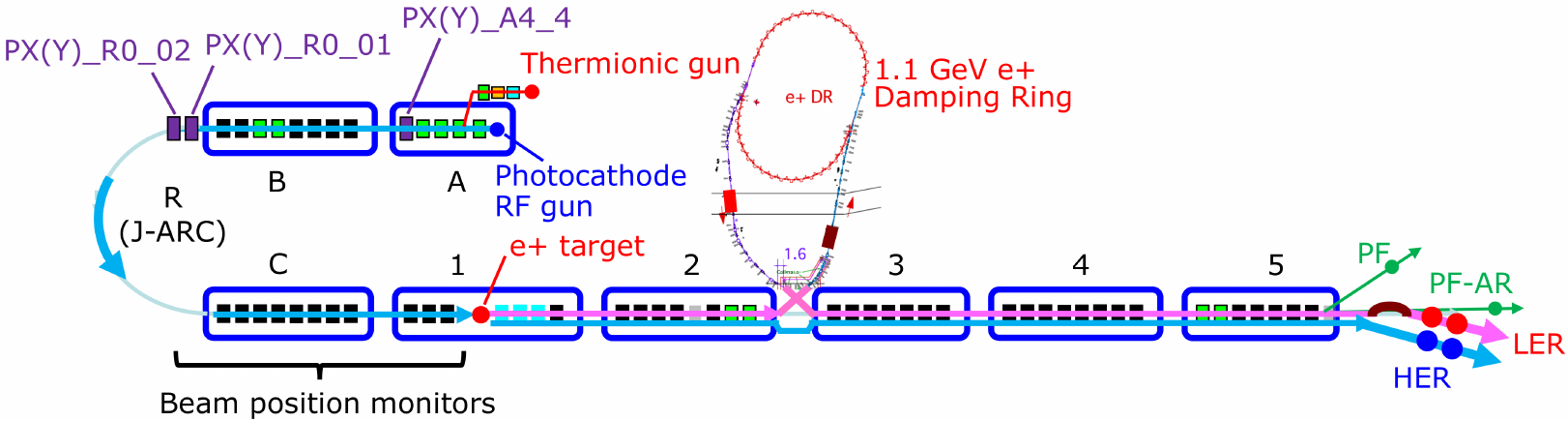}
 \caption{Layout of pulsed steering magnets and beam position monitors at KEK Linac.}
 \label{fig:linac_layout}
\end{figure*}

This section describes the properties and layout of the pulsed steering magnets
and beam position monitors (BPMs) used in the beam-tuning experiments at the KEK
Linac.

Figure~\ref{fig:linac_layout} shows a schematic illustration of the Linac.
First, an electron beam is generated using electron guns, i.e., a thermionic DC
gun and a photocathode RF gun~\cite{SuperKEKB_TDR}.
The electron beam used to generate positrons in the beam-tuning experiments,
which is known as the KBP beam, is generated by a thermionic DC gun suitable for
generating high-charge beams. After passing through the
bunchers~\cite{SuperKEKB_TDR}, the electron beam enters the straight A--B
sectors, followed by the arc sector known as the R sector.
After turning around in the R sector, the \SI{1.5}{\giga\electronvolt} electron
beam entering the C sector is further accelerated to
\SI{3.3}{\giga\electronvolt} and strikes a positron-generating target composed
of tungsten in sector 1~\cite{kamitani:ipac14-mopri004}. Positrons generated via
multiple scattering in the tungsten target and the subsequent electron-positron
pair generation are focused forward by the flux
concentrator~\cite{enomoto:ipac2021-wepab144}.
The positron beam is accelerated to \SI{1.1}{\giga\electronvolt} and injected
into the damping ring~\cite{Iida:eeFACT2018-TUPAB07} from sector 2.
During the \SI{40}{\milli\second} storage in the damping ring, the emittance is
reduced via radiation damping. In the switchyard downstream of sector 5, the
positron beam is injected into the beam transport line, thus resulting in a
SuperKEKB positron ring~\cite{SuperKEKB_TDR} (denoted as LER in
Fig.~\ref{fig:linac_layout}).

The beam-tuning experiment for single-objective optimization aims to optimize
the current applied to the coils of the pulsed steering
magnet~\cite{Enomoto:LINAC2018-WE1A06} and consequently maximize the
electron-beam charge arriving at the tungsten target.
Such charge-maximization tunings are typically performed manually by operation
experts. The current experiment was conducted to determine whether an
optimization program can replace expert-based tunings.
Six pulsed steering magnets were used in the beam-tuning experiments. Two of the
pulsed steering magnets were PX(Y)\_A4\_4 (X is horizontal, Y is vertical) at
the A-sector end, and four were PX(Y)\_R0\_01 and PX(Y)\_R0\_02 near the
entrance of the R sector, thus totaling six steering magnets.
PX(Y)\_A4\_4, PX(Y)\_R0\_01 and PX(Y)\_R0\_02 are six pulsed steering magnets
installed from the A-sector end to the R-sector end. A total of 14
BPMs~\cite{Satoh:IPAC2018-THPMF045, miyahara:ibic2019-wepp006} for the
beam-charge measurements, each with a resolution of approximately
\SI{1}{\percent}, were selected from downstream of the R sector to immediately
before the tungsten target.

We set the applied current and acquired the measured charge using the EPICS
protocol~\cite{DALESIO1994179}. Because the electron-beam repetition rate during
the beam experiment was \SI{1}{\hertz}, a wait time of \SI{1}{\second} was
allowed after changing the applied current until the change in the applied
current was reflected in the beam-orbit modification.
The measured charges were averaged for all BPMs every second, and the operation
was repeated two more times and averaged (which required \SI{3}{\second}) to
obtain a better charge-measurement resolution.

\subsection{Experimental results}\label{sec:result1}

This section presents the results of the beam-tuning experiments conducted at
the Linac in June 2023.

\begin{figure}[hbt]
 \centering
 \includegraphics[width=0.49\columnwidth]{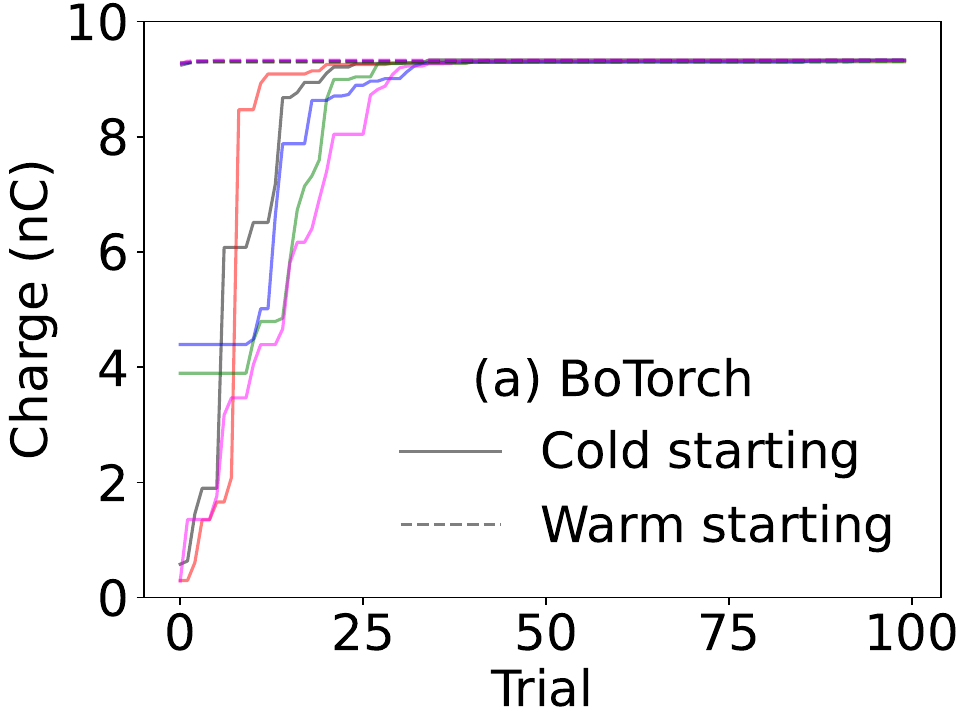}
 \includegraphics[width=0.49\columnwidth]{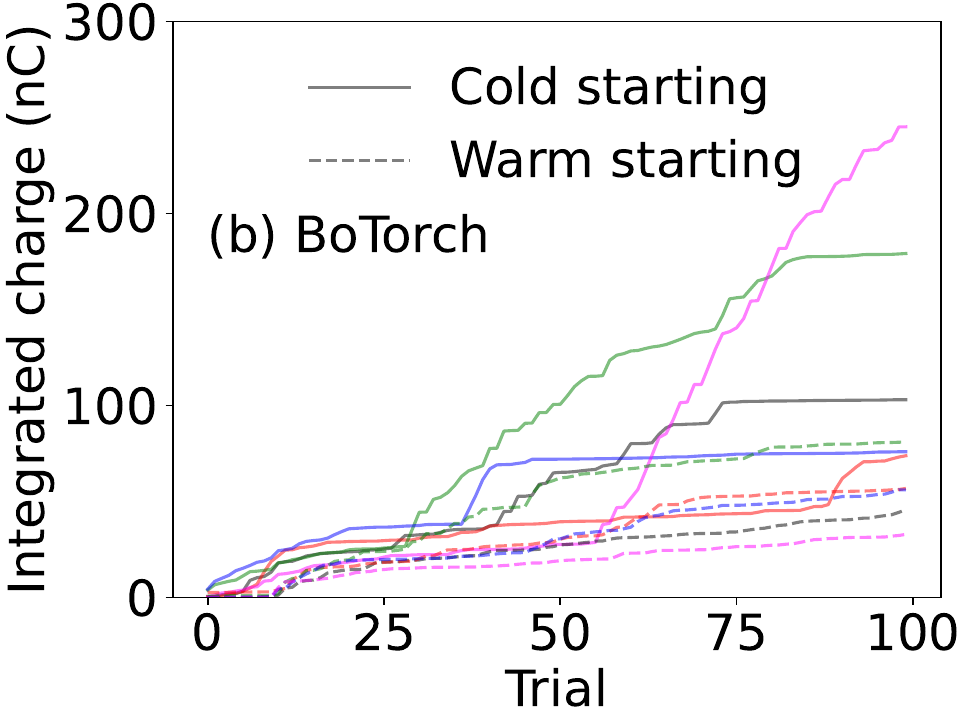}
 \includegraphics[width=0.49\columnwidth]{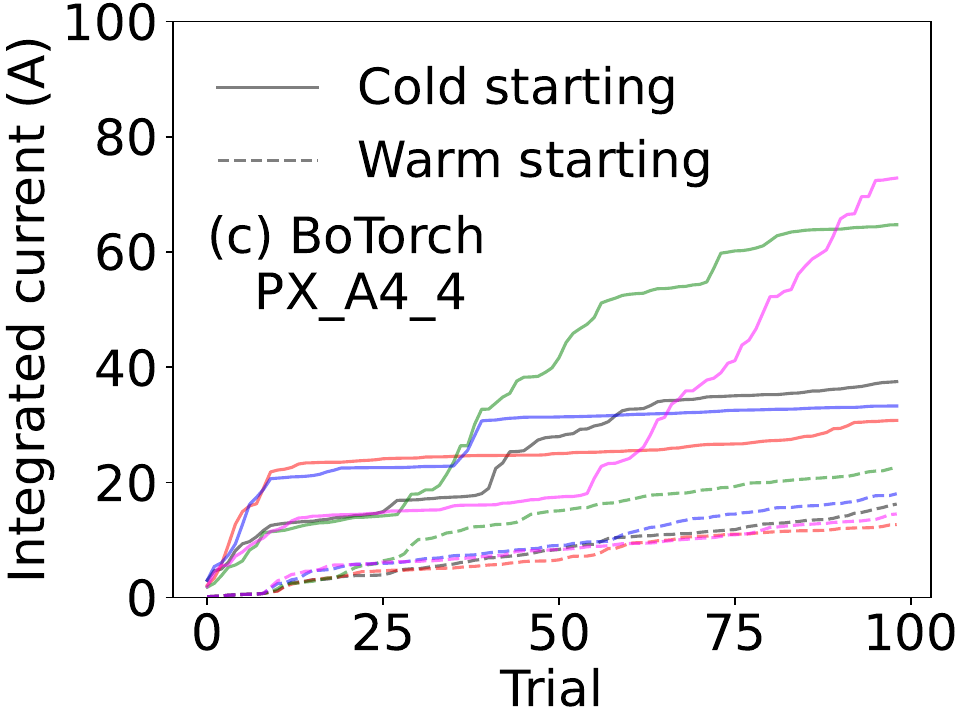}
 \includegraphics[width=0.49\columnwidth]{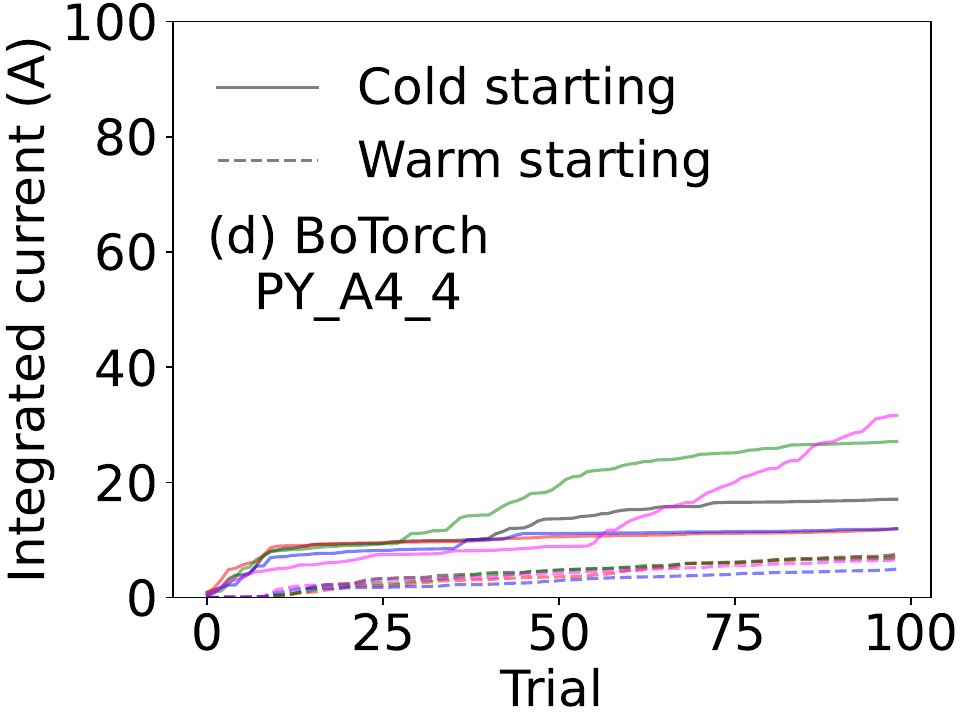}
 \caption{Results obtained using Bayesian optimization based on BoTorch algorithm. Panel
 (a) indicates peak hold values of electron-beam charge. Panel (b) indicates
 integrated charge over trials. Panels (c) and (d) indicate integral applied
 current for PX\_A4\_4 and PY\_A4\_4, respectively.}
 \label{fig:BoTorch}
\end{figure}

Figure~\ref{fig:BoTorch} shows the results obtained using Bayesian optimization
based on the BoTorch algorithm. Panel (a) shows the peak hold values of the
electron-beam charge, which varied from the 1st to the 100th trial.
The five solid lines (each referred to as a run) represent the cases in which
the optimization parameters and applied currents of the coils are set randomly
within the configuration domain during initialization. Hereafter, these runs are
referred to as \textit{``cold starting.''} The initialization was immediately
interleaved after the optimization program was started.
Initialization was performed 10 times for the first to tenth trials, which was
necessary to obtain an estimation of the probability density distribution over
the configuration domain. One run for every 100 trials required approximately 20
min, which constituted primarily the wait time required to average the charge
information from the BPMs.
In two of the five runs, a charge of approximately \SI{4}{\nano\coulomb} was
obtained in the first trial immediately after initialization. In all five runs,
a maximum electron-beam charge exceeding \SI{9.3}{\nano\coulomb} was reached in
approximately 35 trials (which required 7 min), which was comparable to the
pre-experiment manual-adjustment results by the experts.

The five dashed lines (which cannot be distinguished easily because they almost
overlap) represent the case in which a combination of applied currents known in
advance to provide a high beam charge is enqueued as the initial configuration.
Hereafter, we refer to these runs as \textit{``warm
starting''}~\cite{Nomura_Watanabe_Akimoto_Ozaki_Onishi_2021}.
The enqueued applied current combinations were copied from the 10 combinations
with the highest-ranking beam charges extracted from one of the cold starting
runs (solid lines).
Consequently, the dashed lines indicate a high peak charge immediately after
initialization. The warm-starting method should benefit actual accelerator
tuning provided that a certain degree of reproducibility can be guaranteed.

Panel (b) shows the integrated charge over trials, which is expressed as
\begin{equation}
Q_\textrm{int}(t) = \sum_{i=2}^t \lvert q(i) - q(i-1) \rvert,
\end{equation}
where $q(i)$ denotes the beam charge obtained in the $i$-th trial. Similar to
panel (b), panels (c) and (d) show the integral applied current for PX\_A4\_4
and PY\_A4\_4, respectively, which is expressed as
\begin{equation}
I_\textrm{int}(t) = \sum_{i=2}^t \lvert I(i) - I(i-1) \rvert.
\end{equation}
Here, $I(i)$ indicates the current applied in the $i$-th trial. The integrated
charge and applied current determine the trial at which the beam charge becomes
steeper or milder. The five solid lines in panels (b)--(d) correspond to the
cold-starting runs, where three ascended considerably from approximately the
30th to the 40th trial. Comparing the three solid lines in panel (a), we
discovered that the optimization shifted from exploitation to exploration around
the trial when the beam charge reached its maximum. Similarly, the solid magenta
line in panel (a), which reached the maximum beam charge the slowest, indicated
a significant shift to exploration at approximately the 55th trial, as shown in
panels (b)--(d). In the five dashed lines for warm starting in panel (b), the
increase in the integrated charge remained gradual after the 20th trial, which
is consistent with the fact that the BoTorch algorithm continued to exploit the
near-optimum applied current, as shown in panels (c) and (d).

\begin{figure}[hbt]
 \centering
 \includegraphics[width=0.49\columnwidth]{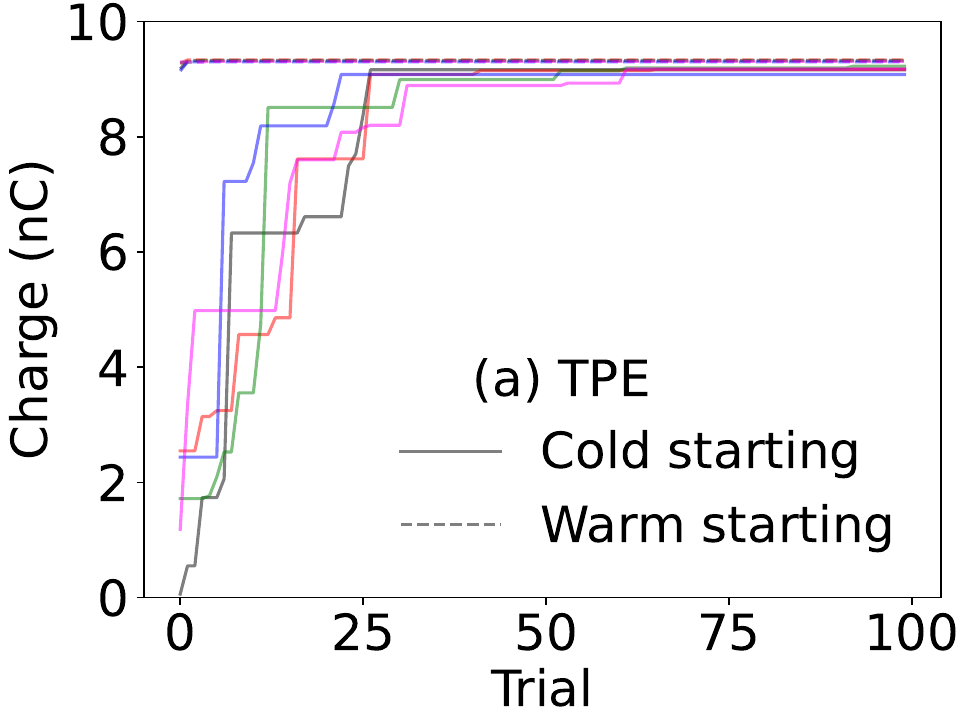}
 \includegraphics[width=0.49\columnwidth]{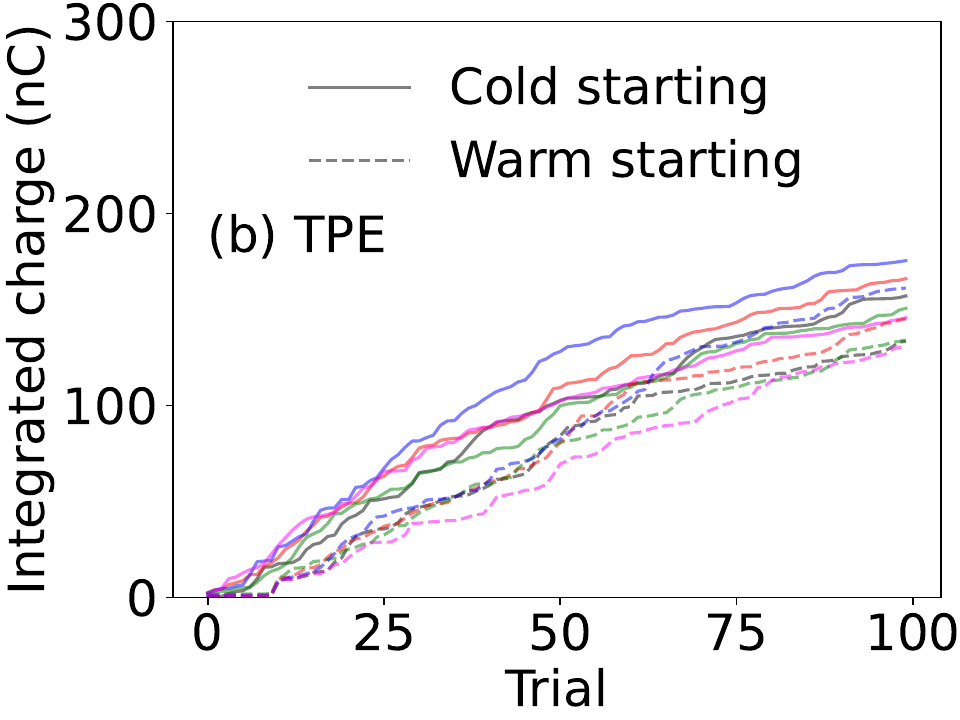}
 \includegraphics[width=0.49\columnwidth]{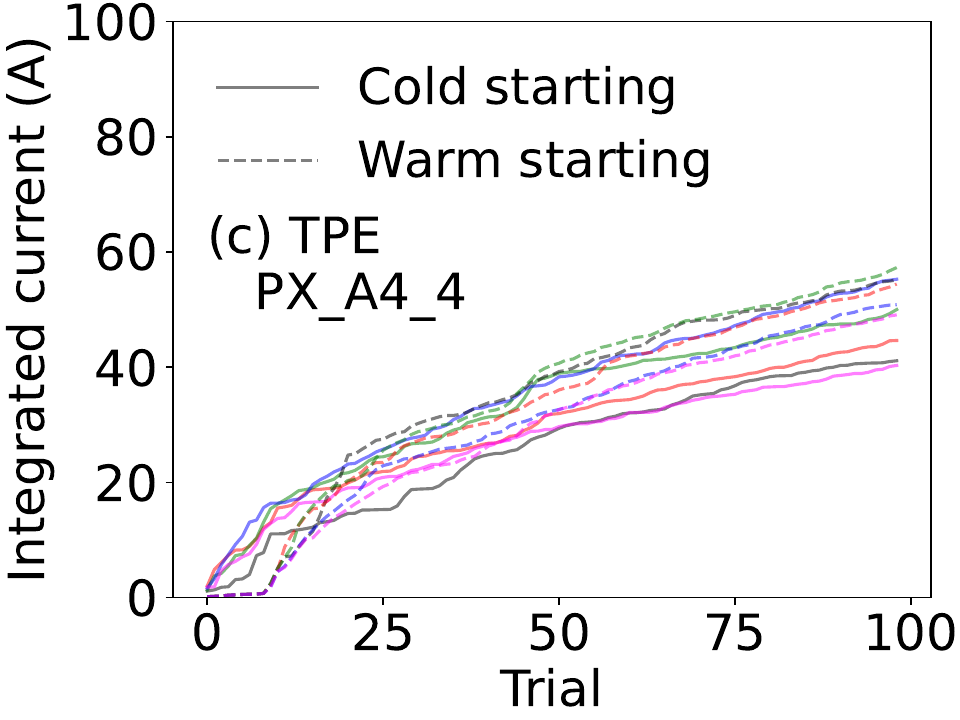}
 \includegraphics[width=0.49\columnwidth]{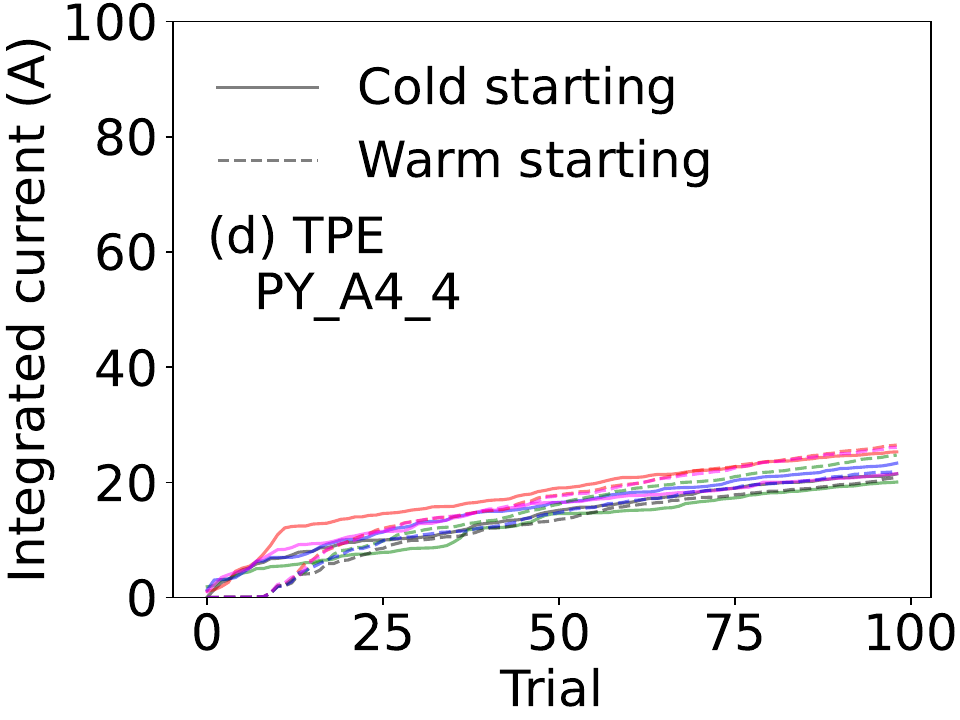}
 \caption{Results obtained using TPE algorithm. Panel (a) indicates peak hold values of
 electron-beam charge. Panel (b) indicates integrated charge over trials. Panels
 (c) and (d) indicate integral applied current for PX\_A4\_4 and PY\_A4\_4,
 respectively.}
 \label{fig:TPE}
\end{figure}

Figure~\ref{fig:TPE} shows the results of the TPE algorithm. We performed a
beam-tuning experiment using BoTorch on June 2, 2023, 11 am--3 pm, and an
experiment using the TPE on the same day, 3--6 pm; the change in the Linac
conditions between the two was negligible.

The five runs indicated by the solid line in panel (a) reached
\SI{9}{\nano\coulomb} around the 40th trial and the maximum charge around the
60th trial. The maximum charge was slightly lower than \SI{9.2}{\nano\coulomb},
which was in fact \SI{0.1}{\nano\coulomb} lower than the result achieved by
BoTorch (see Fig.~\ref{fig:BoTorch} (a)).
The dashed lines in panel (a) show the cases of ``warm starting.'' Ten
combinations of the applied currents were obtained from the cold-starting
BoTorch results shown in Fig.~\ref{fig:BoTorch} (a). As expected from the
superior combinations, the maximum charge of the warm-starting runs in
Fig.~\ref{fig:TPE} (a) exceeded \SI{9.3}{\nano\coulomb}.

Panel (b) shows that the integrated charge obtained by the cold-starting TPE was
generally higher than that of BoTorch, as shown in Fig.~\ref{fig:BoTorch} (b),
regardless of whether cold or warm starting was used.
Based on panels (b)--(d), the TPE algorithm might focus more on exploration than
exploitation and thus reach its maximum charge slower than BoTorch. In panel
(b), we observed that up to the 50th trial, the integrated charge for the
warm-starting runs was slightly smaller than that for cold-starting runs;
however, after the 50th trial, the difference between the two became less
prominent.
This transition may be because, as shown in panels (c) and (d), the
warm-starting TPE algorithm searches a broader range of configuration parameter
domains than the cold-starting TPE after initialization. In panels (c) and (d),
we observed that the integral applied current of warm starting exceeded that of
cold starting around the 30--40th trials.

\begin{figure}[hbt]
 \centering
 \includegraphics[width=0.49\columnwidth]{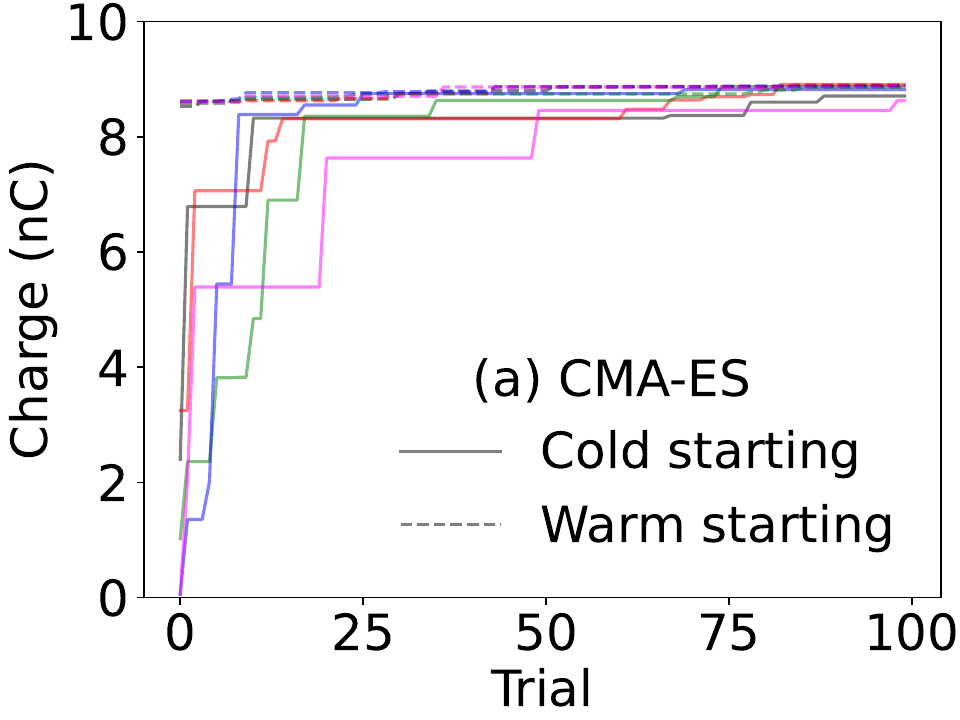}
 \includegraphics[width=0.49\columnwidth]{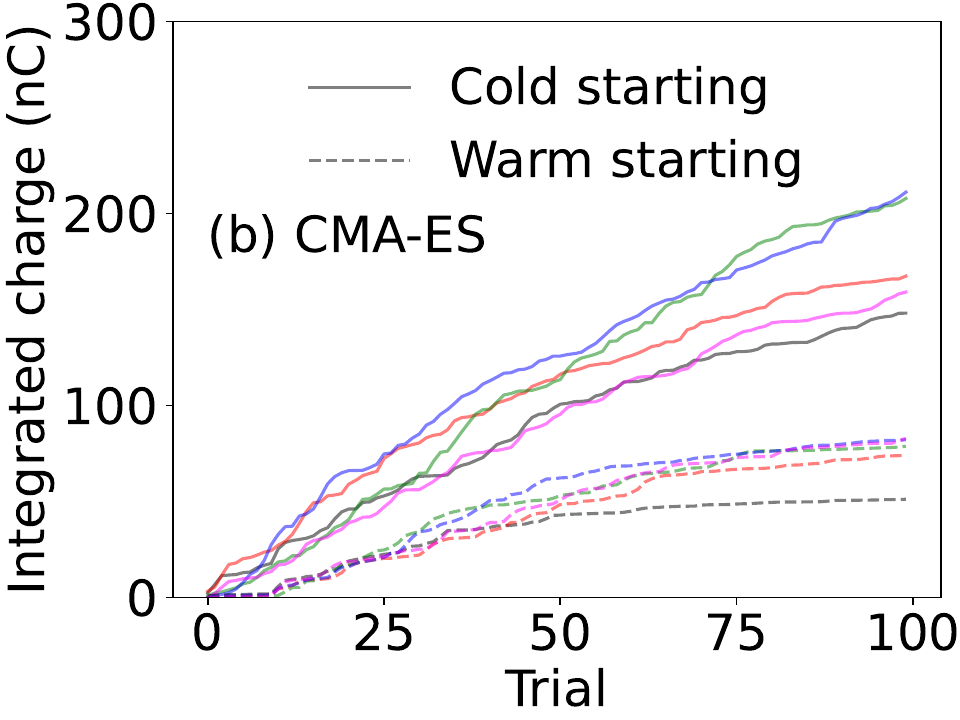}
 \includegraphics[width=0.49\columnwidth]{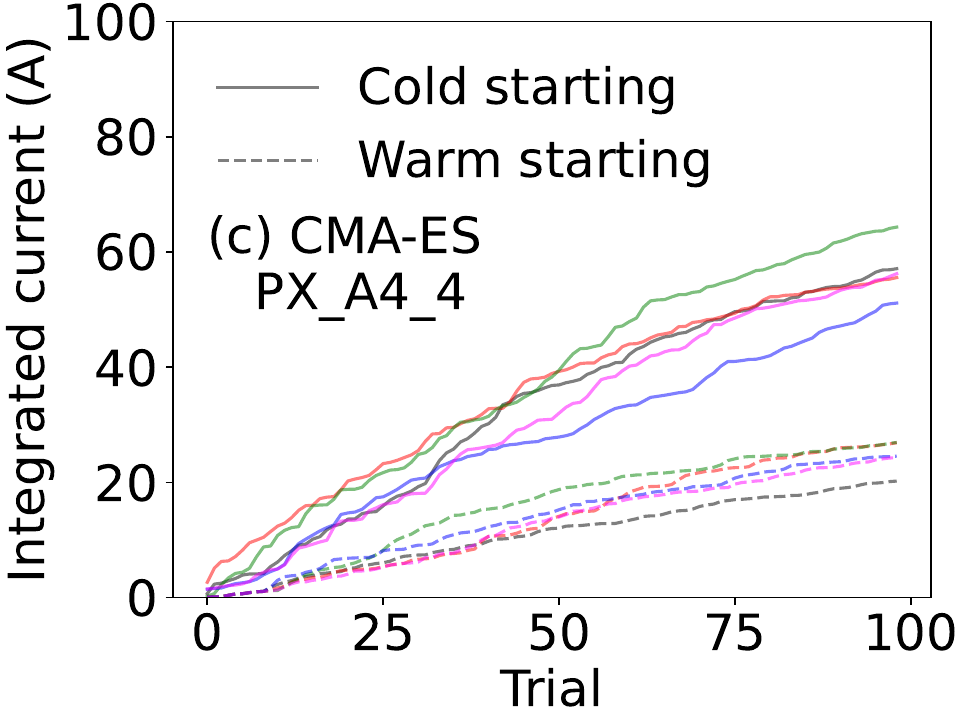}
 \includegraphics[width=0.49\columnwidth]{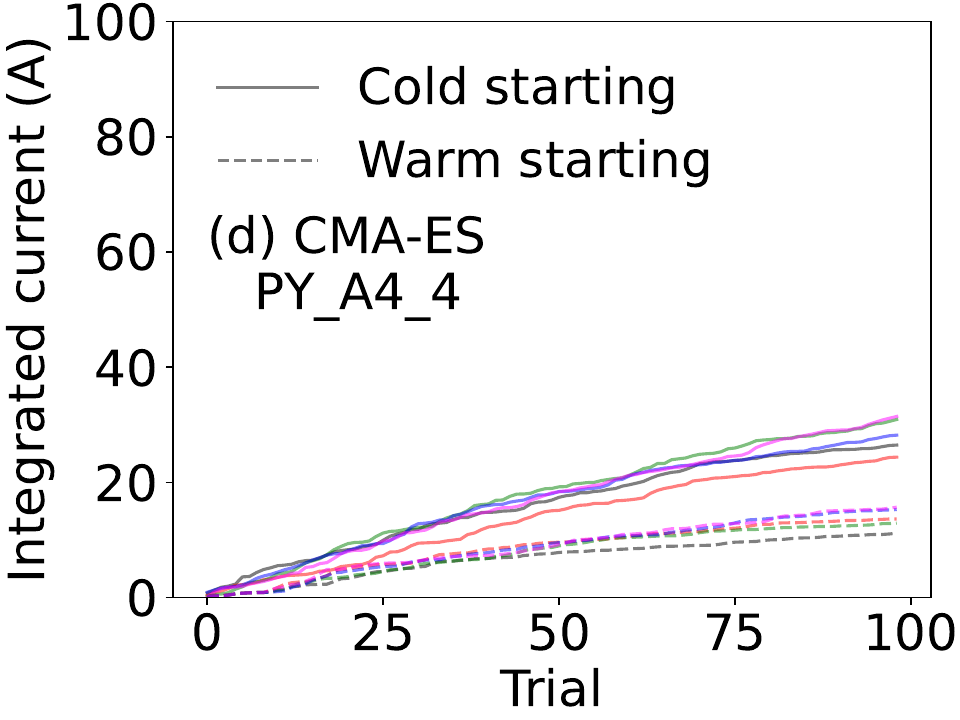}
 \caption{Results obtained using CMA-ES algorithm. Panel (a) indicates peak hold values
 of electron-beam charge. Panel (b) indicates integrated charge over trials.
 Panels (c) and (d) show integrated applied currents for PX\_A4\_4 and
 PY\_A4\_4, respectively.}
 \label{fig:CMAES}
\end{figure}

Figure~\ref{fig:CMAES} shows the experimental results obtained using the CMA-ES.
The measurements were performed on June 12, i.e., 10 days after the BoTorch
measurements, as shown in Fig.~\ref{fig:BoTorch}, and the TPE measurements are
shown in Fig.~\ref{fig:TPE}. During those 10 days, the situation downstream of
the electron-beam direction changed, and the beam charge arriving at the most
upstream pulsed steering magnet, PX(Y)\_A4\_4, decreased. Therefore, the maximum
charge obtained via the CMA-ES was lower, i.e., below \SI{9}{\nano\coulomb}.
In panel (a), the solid lines for the cold-starting case shows the maximum
charge increasing the most slowly among the three algorithms considered. The
dashed lines show the warm-starting case, where the 10 combinations of the
applied currents were obtained from the cold-starting CMA-ES measurements (i.e.,
one of the solid lines).
As shown in panels (b)--(d), the integrated charge and applied current were
consistently lower for warm starting than for cold starting. This tendency
indicates that, as with BoTorch, in the case of warm starting, the optimized
applied current is enqueued as the initial value combination and the surrounding
domain of that combination is exploited.

The three algorithms, shown in Figs.~\ref{fig:BoTorch}--\ref{fig:CMAES}, were
terminated after 100 trials.
However, for BoTorch and the TPE, only one run each was tested, and the
optimization was extended to 300 trials. Based on the results, BoTorch continued
to be optimized, thus emphasizing exploitation near the maximum charge. However,
the TPE shifted from exploitation to exploration after the 150th trial.
Systematic measurements based on a significantly higher number of runs shall be
attempted in future studies.

\begin{figure}[hbt]
 \centering
 \includegraphics[width=0.49\columnwidth]{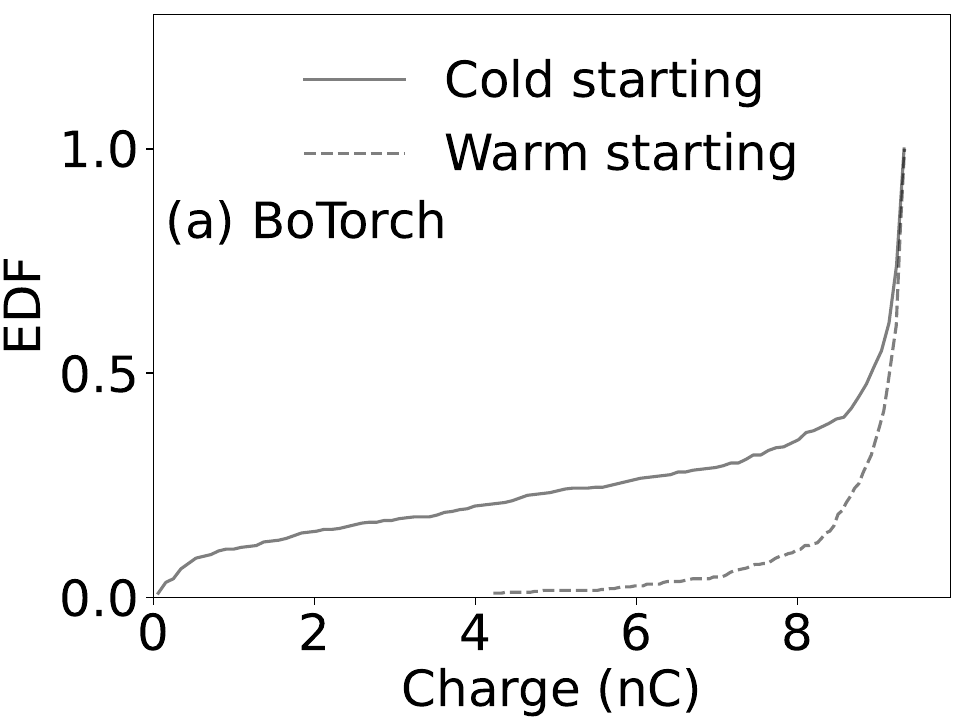}
 \includegraphics[width=0.49\columnwidth]{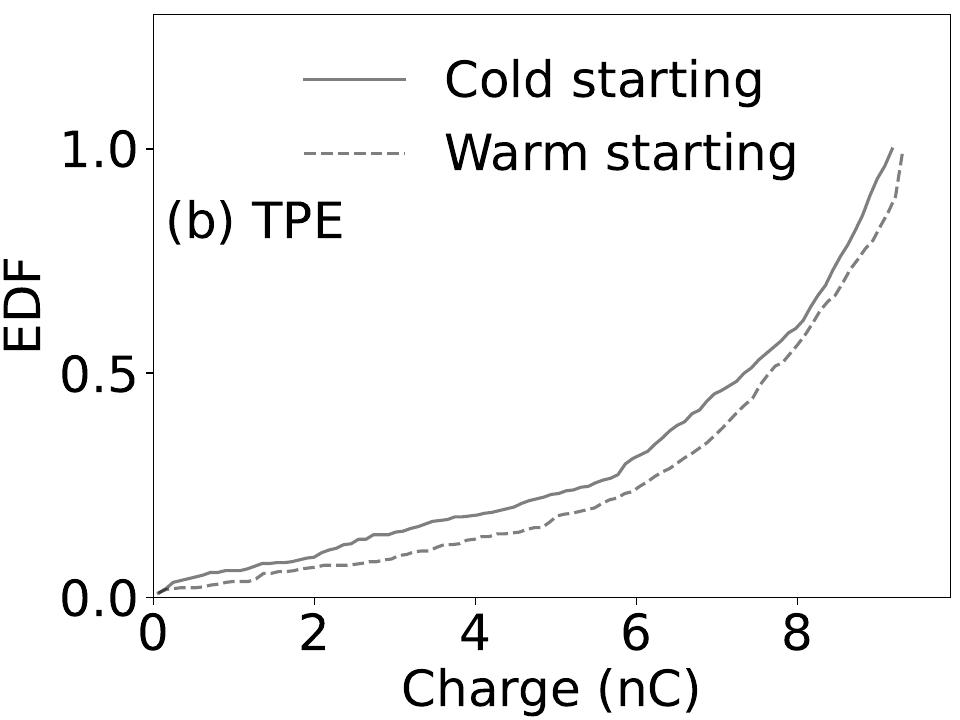}
 \includegraphics[width=0.49\columnwidth,left]{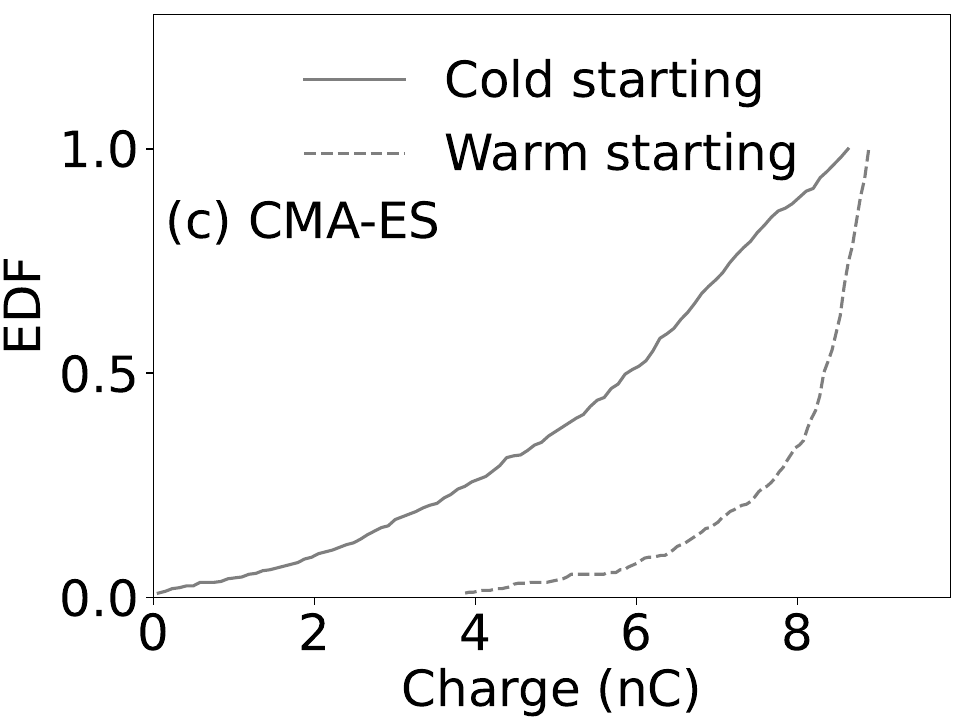}
 \caption{Empirical distribution functions (EDF) averaged over five runs. Panels (a),
 (b), and (c) show results of BoTorch, TPE, and CMA-ES, respectively.}
 \label{fig:EDF}
\end{figure}

Figure~\ref{fig:EDF} shows the empirical distribution function (EDF), which is
defined as
\begin{equation}
F(q_\textrm{thr}) = \frac{\textrm{number of trials of } q < q_\textrm{thr}}{100} = \frac{1}{100}\sum_{i=1}^{100}\mathds{1}_{q < q_\textrm{thr}}.
\end{equation}
Panel (a) shows the results of the BoTorch algorithm averaged over five runs for
the cold-starting (solid line) and warm-starting (dashed line) cases. Each line
shows a steep increase in the EDF beginning at approximately
\SI{8.5}{\nano\coulomb}, thus indicating that many trials yielded a beam charge
exceeding \SI{8.5}{\nano\coulomb}. For cold starting, approximately
\SI{60}{\percent} of the trials were distributed above \SI{8.5}{\nano\coulomb},
whereas for warm starting, approximately \SI{80}{\percent} of the trials were
distributed above \SI{8.5}{\nano\coulomb}.
The results shown in panel (a) are consistent with the superior performance of
BoTorch shown in Fig..~\ref{fig:BoTorch}.

Panel (b) shows the results of the TPE measured on the same day as that of
BoTorch. For a specified beam charge, the warm-starting TPE (dashed line)
consistently showed a lower EDF than the cold-starting TPE (solid line), thus
indicating that the warm-starting trials were distributed at a slightly higher
beam charge. However, the shapes of the EDFs were almost identical, and the
differences were insignificant compared with those of BoTorch and the CMA-ES.
Slight differences between the two curves suggest that the warm-starting TPE
affects the maximum charge through initialization but negligibly affects the
trade-off between exploration and exploitation. This trend is consistent with
the integrated charge and applied current distributions shown in
Figs.~\ref{fig:TPE} (b)--(d).

Panel (c) shows the measurement results obtained using the CMA-ES, where the EDF
integrated from 0 to \SI{8}{\nano\coulomb} for cold starting was the largest
among the three algorithms considered. Similar to the results shown in
Fig.~\ref{fig:CMAES} (a), the CMA-ES algorithm for cold starting performed an
exploration-oriented optimization in this experiment. However, for warm
starting, more than \SI{70}{\percent} of the trials occurred at
\SI{8}{\nano\coulomb} or higher, thus indicating that the optimization focused
on exploitation. The trends in panel (c) are similarly shown in
Figs.~\ref{fig:CMAES} (b)--(d) for warm starting (dashed line).

\begin{figure}[hbt]
 \centering
 \includegraphics[width=0.49\columnwidth]{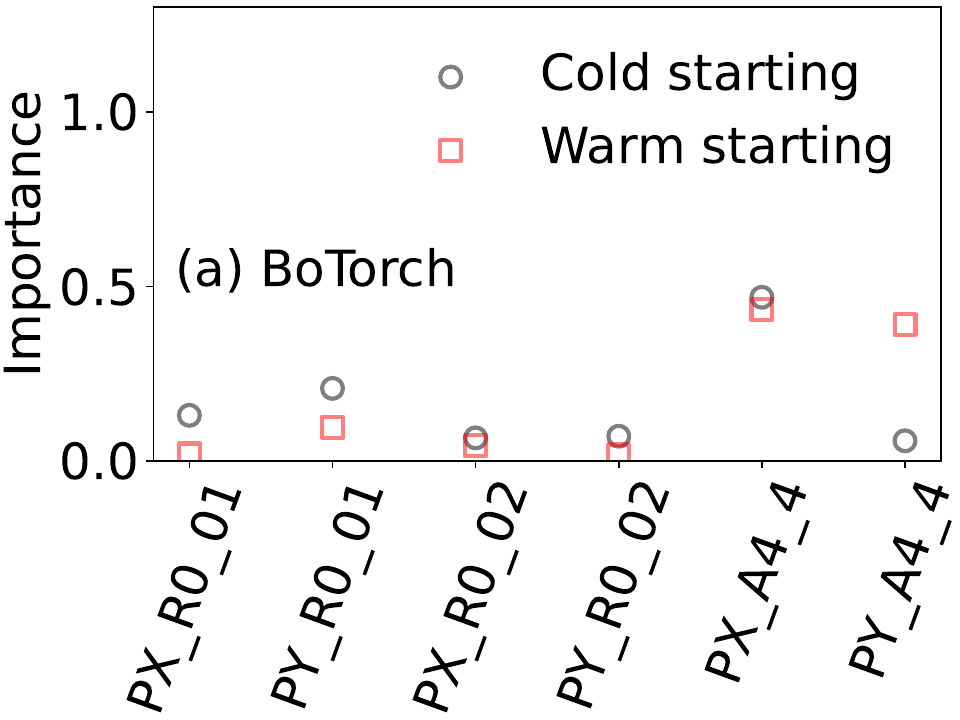}
 \includegraphics[width=0.49\columnwidth]{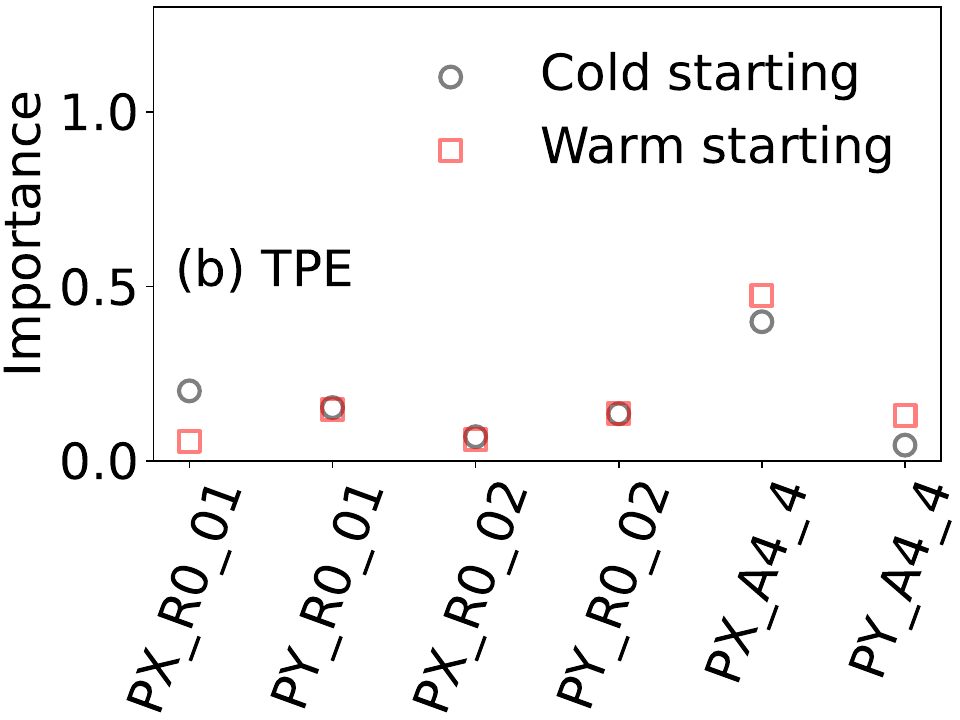}
 \includegraphics[width=0.49\columnwidth,left]{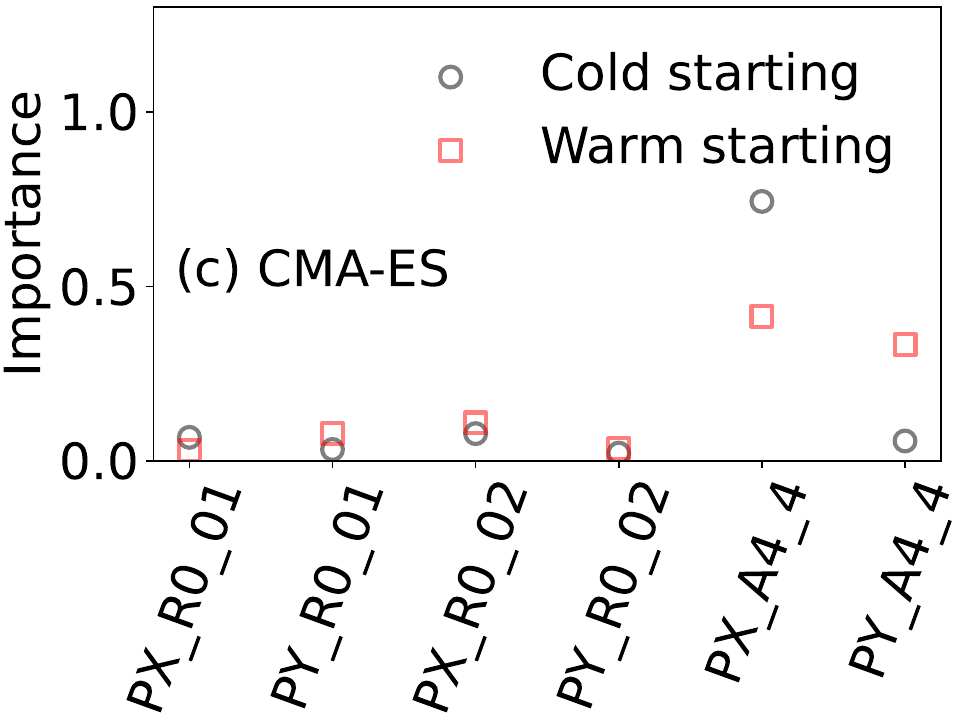}
 \caption{Importance of each input parameter averaged over five runs. Panels (a), (b),
 and (c) show results of BoTorch, TPE, and CMA-ES, respectively.}
 \label{fig:Importance}
\end{figure}

Figure~\ref{fig:Importance} shows the importance of each parameter, namely, the
effectiveness of the applied currents in increasing the beam charge. The
importance of the parameters can be quantified using a method proposed in
Ref.~\cite{pmlr-v32-hutter14}, which is based on a random forest prediction
model~\cite{10.1007/978-3-642-25566-3_40} and functional analysis of variance
(fANOVA). First, a random-forest model was established to predict the average
algorithm performance over the configuration domain. Subsequently, the fANOVA
decomposes the variance of the overall algorithm performance into additive
components, with each corresponding to a subset of the algorithm parameters.
Finally, the fraction of variance associated with each subset of parameters
relative to the overall performance variance quantifies the importance of the
corresponding subset.

Panel (a) shows the results of the BoTorch measurements.
The black circles and red squares represent the average of five cold- and
warm-starting runs, respectively.
For cold starting, the maximum importance was $\sim0.5$ for pulsed steering
magnet PX\_A4\_4.
Because the importance was normalized such that the sum was 1, PX\_A4\_4 alone
appeared to have contributed to approximately \SI{50}{\percent} of the
importance.
As shown in Fig.~\ref{fig:linac_layout}, PX\_A4\_4 was the most upstream of the
three horizontal magnets used in the beam-tuning experiment.
Therefore, PX\_A4\_4 is expected to exert the most significant effect on the
horizontal orbit modification, thus resulting in a higher importance for the
objective function (i.e., beam charge).
Because of a sizable orbit error in the horizontal direction, the beam
established contact with the beam collimator in the arc R sector, thus resulting
in a significant beam-charge loss. This may explain the higher importance of the
horizontal direction PX\_A4\_4 compared with that of the vertical direction
PY\_A4\_4.
For warm starting, the importance of PY\_A4\_4 increased to 0.4. As shown by the
integrated charge in Fig.~\ref{fig:BoTorch} (b), the integrated applied current
in Figs.~\ref{fig:BoTorch} (c) and (d), and the EDF in Fig.~\ref{fig:EDF} (a),
most of the 100 parameter sets for warm starting did not change significantly
from the best parameter set that yielded the maximum charge. That is, the
applied current of PY\_A4\_4, which was located upstream, was almost optimized
immediately after the initialization. Thus, a slight change in the applied
current significantly affected the beam charge loss and became more important.
In two locations in the R sector, the vertical beta function exceeded
\SI{100}{\meter}. In addition, an electron beam generated by the thermionic DC
gun indicated a large emittance. Therefore, if the beam orbit is shifted
vertically, then a portion of the bunch with a large transverse size hits the
beam pipe, thus resulting in beam-charge loss.

Panel (b) presents the TPE results. The results of cold and warm starting were
similar.
As shown in Figs.~\ref{fig:TPE} (b)--(d) and Fig.~\ref{fig:EDF} (b), the
trade-off between exploitation and exploration changed only slightly, regardless
of whether cold or warm starting was used in TPE.
Thus, we can assume that the importance of each parameter is similarly
distributed for both the cold- and warm-starting runs. Because the TPE optimizes
with emphasis on exploration even for warm starting, the results achieved are
comparable to those yielded by BoTorch for cold starting.

Panel (c) presents the CMA-ES results.
The results for warm starting were comparable to those of BoTorch, which is as
expected owing to the similarity of the EDFs shown in Figs.~\ref{fig:EDF} (a)
and (c). Meanwhile, the importance of PX\_A4\_4 for cold starting exceeded 0.7
for the CMA-ES, as compared with 0.5 for BoTorch.
Although we have yet to achieve quantitative understanding, we hypothesize that
the anti-correlation between the more critical PX\_A4\_4 and less critical
PX\_R0\_01 may affect the importance of PX\_A4\_4. Based on panels (a) and (b),
the importance of PX\_R0\_01 is 0.15--0.2, whereas it is less than 0.1 in panel
(c).

\begin{figure}[hbt]
 \centering
 \includegraphics[width=\columnwidth]{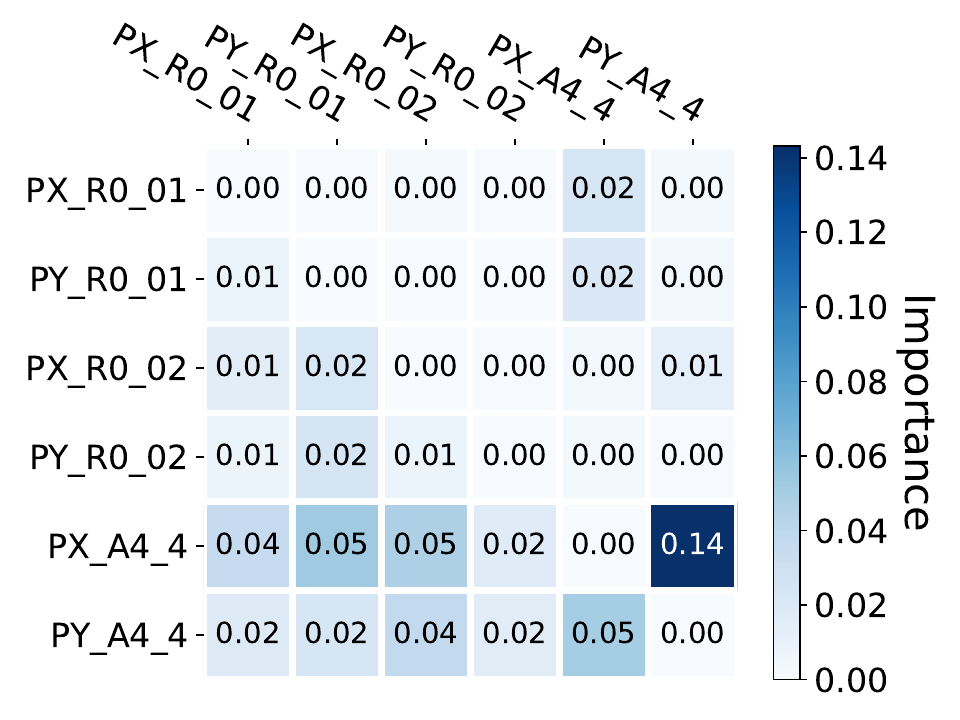}
 \caption{Importance of each input parameter combination. Numbers shown above and below
 diagonal line indicate BoTorch and TPE results, respectively. Both results are
 for warm starting.}
 \label{fig:Importance_corr}
\end{figure}

Figure~\ref{fig:Importance_corr} shows the importance of the two-parameter
combinations. The results yielded by the warm-starting BoTorch and warm-starting
TPE are shown above and below the diagonal line, respectively. For simplicity,
we set the values on the diagonal line to zero.
In the BoTorch results, the combination of PX\_A4\_4 and PY\_A4\_4 showed the
highest importance of $\sim0.14$. Combining the results for warm starting (black
squares) in Fig.~\ref{fig:Importance} (a), we observed that the order of
importance from highest to lowest was PX\_A4\_4, PY\_A4\_4, the combination of
both, and the other parameters.
 
Meanwhile, the importance of the other pulsed steering magnets and their
combinations was less than 0.1.
The TPE results below the diagonal line show that the importance for the
combination of each parameter was 0.05 at the maximum. PY\_A4\_4 was less
critical in the TPE, even for warm starting, as shown in
Fig.~\ref{fig:Importance} (b). Consequently, the combination of PX\_A4\_4 and
PY\_A4\_4 was less critical.

Figures\ref{fig:Importance} and \ref{fig:Importance_corr} show that the applied
currents of PX\_A4\_4 and PY\_A4\_4, which were the most upstream pulsed
steering magnets used in the experiment, functioned similarly as the other
steering magnets for the three algorithms. Therefore, only the applied currents
of PX\_A4\_4 and PY\_A4\_4 appeared to be sufficient for the enqueued initial
parameter values considered under warm starting.

In the next beam-tuning experiment, we plan to perform beam tuning at different
sectors of Linac using other magnets and other types of objective functions to
obtain more general insights into machine learning-assisted beam tuning.

\section{Multiobjective optimization}\label{sec:multi}

This section provides an overview of the methodology of multiobjective
optimization, i.e., the case in which multiple objective functions are optimized
simultaneously. The $M$-dimensional multiple-objective functions are denoted as
$f^{(1)}, f^{(1)}, \ldots, f^{(M)}$.
In the beam-tuning experiment, we set $M=2$ to accommodate the two objective
functions (maximization of the beam charge and minimization of the dispersion
function).
For simplicity, we assume that the goal of multiobjective optimization is to
maximize all dimensions.

\begin{figure}[hbt]
 \centering
 \includegraphics[width=\columnwidth]{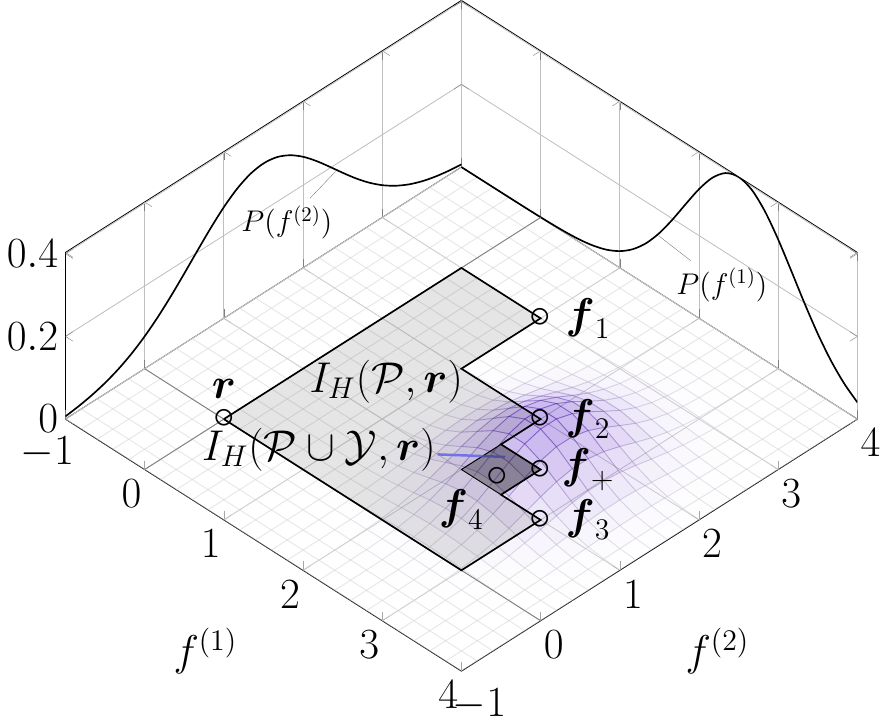}
 \caption{Example of hypervolume indicator $I_H$ with Pareto solution set $\{(\boldsymbol{x}_i,\boldsymbol{f}_i)\}_{i=1}^3$.}
 \label{fig:MOBO}
\end{figure}

The simultaneous optimization of multiobjective functions can be redefined as
obtaining all the Pareto optimal solutions.
To illustrate the Pareto optimal solution, we define the dominance relation. For
the two objective functions $\boldsymbol{f}(\boldsymbol{x})$ and
$\boldsymbol{f}(\boldsymbol{x'})$, the relation
\begin{equation}
	\boldsymbol{f}(\boldsymbol{x}) \succeq \boldsymbol{f}(\boldsymbol{x'})
	\Leftrightarrow \forall m \in \{1,\ldots,M\} f^{(m)}(\boldsymbol{x}) \geqq f^{(m)}(\boldsymbol{x'})
	\label{eq:dominance-relation}
\end{equation}
indicates that $\boldsymbol{f}(\boldsymbol{x})$ dominates
$\boldsymbol{f}(\boldsymbol{x'})$ if $f^{(m)}(\boldsymbol{x})$ is greater than
or equal to $f^{(m)}(\boldsymbol{x'})$ for all dimensions. Here,
$\boldsymbol{x}_i$ represent the input variables (applied currents of the pulsed
steering magnets) and $\boldsymbol{f}_i$ the objective functions (a beam charge
and a dispersion function).
The last inequality is presented in Eq.~(\ref{eq:dominance-relation}) as we are
addressing a maximization problem.
For example, as shown in Fig.~\ref{fig:MOBO}, $\boldsymbol{f}_+$ dominates
$\boldsymbol{f}_4$.
We regard $\boldsymbol{f}$ as a Pareto-optimal solution when no other point in
the objective function space dominates $\boldsymbol{f}$.
Generally, more than one Pareto-optimal solution exists for multiple objective
functions.

The Pareto front is the surface created when plotting the Pareto solution set.
Multiobjective optimization aims to efficiently obtain many Pareto-optimal
solutions near the Pareto front by discounting the superiority or inferiority of
the multiple Pareto-optimal solutions.
To obtain the Pareto front, we introduce a Pareto hypervolume. Let $\mathcal{D}
= \{(\boldsymbol{x}_i, \boldsymbol{f}_i)\}_{i=1}^N$ denote the current dataset,
where $N$ is the dataset size. The Pareto front in the data set $\mathcal{D}$
expands with each data addition.

Once a reference point is determined to evaluate the expansion, a hyperrectangle
can be defined using the reference point $\boldsymbol{r}$ and the Pareto
solution set $\mathcal{P} \in \mathcal{D}$.
\begin{equation}
	\cup_{i=1}^{\mathcal{P}}[\boldsymbol{r}, \boldsymbol{f}_i].
	\label{eq:rectangle}
\end{equation}
The hypervolume indicator in Eq.~(\ref{eq:rectangle}) is the $M$-dimensional
Lebesgue measure, which is expressed as
\begin{equation}
	I_H(\mathcal{P}, \boldsymbol{r}) = \lambda_M(\cup_{i=1}^{\mathcal{P}}[\boldsymbol{r}, \boldsymbol{f}_i]).
	\label{eq:IH}
\end{equation}
The shaded light-gray area in Fig.~\ref{fig:MOBO} shows the hypervolume
indicator with the Pareto solution set $\mathcal{P} =
\{(\boldsymbol{x}_i,\boldsymbol{f}_i)\}_{i=1}^3$, and reference point
$\boldsymbol{r}$. The hypervolume indicator increases monotonically for each
additional data point.
The shaded dark-gray area shows an increase in the hypervolume indicator owing
to new observations $\mathcal{Y} = (\boldsymbol{x}_+,\boldsymbol{f}_+)$.
Based on this increase, we can define an acquisition function, i.e., the
expected hypervolume improvement (EHVI), which is the expected increment in the
hypervolume indicator before and after obtaining a new observation
$\mathcal{Y}$.
\begin{multline}
	\mathit{EHVI}(\mathcal{Y}, \mathcal{P}, \boldsymbol{r}) = \\
	\int \left( I_H(\mathcal{P}\cup\mathcal{Y}, \boldsymbol{r}) - I_H(\mathcal{P}, \boldsymbol{r}) \right)p(\boldsymbol{f}_+ \mid \boldsymbol{x}_+,\mathcal{D})d\boldsymbol{f}_+.
	\label{eq:EHVI}
\end{multline}
The posterior $p(\boldsymbol{f}_+ \mid \boldsymbol{x}_+,\mathcal{D})$, which is
the distribution indicated by the blue surface in Fig.~\ref{fig:MOBO}, is
approximated via a Gaussian process in the Bayesian optimization.
Analogous to the single-objective optimization, as discussed in
Sec.~\ref{sec:single}, the location $\boldsymbol{x}$ that maximizes the
acquisition function EHVI is selected and then input to the objective function
to advance the Pareto front.
The methodology described above assumes a multiobjective Bayesian optimization.
A detailed description of this multiobjective optimization based on the TPE
algorithm is provided in Refs.~\cite{10.1145/3377930.3389817,
10.1613/jair.1.13188}.

\section{Application to beam charge and dispersion simultaneous optimization}\label{sec:multi-test}

\subsection{Experimental setup at KEK Linac}\label{sec:setup2}

To investigate the feasibility of applying multiobjective Bayesian optimization
to accelerator tuning, we attempted to simultaneously maximize the electron-beam
charge and minimize the dispersion function.

The unexpected dispersion function in the Linac and beam-transport line
increases the emittance of the injection beam and further reduces the injection
efficiency to the light source storage rings (PF and PF-AR) and
SuperKEKB~\cite{SuperKEKB_TDR} downstream of the beam-transport line.
Therefore, the beam must be adjusted such that the sizable dispersion function
in sector R (see Fig.~\ref{fig:linac_layout}) does not leak downstream.

In this beam-tuning experiment, we used an electron beam generated by a
thermionic DC gun dedicated for positron generation. Notably, the dispersion
function in sector 1 does not significantly affect positron generation in the
actual operation. We attempted to minimize the dispersion function in this
experiment such that better electron beams would be received by the PF, PF-AR,
and SuperKEKB. Notably, the electron beam generated by the RF electron gun was
supplied to the PF and PF-AR when this experiment was conducted and was not used
in this study.

In this study, the dispersion function was not used as an objective function.
Instead, the following dispersion-position function was used for simplicity,
which multiplies the square of the dispersion function by the sum of the squares
of the horizontal and vertical positions.
\begin{equation}
f_\textit{disp-pos} =
\sum_{i=1}^{14}(\eta_{x,i}^2+\eta_{y,i}^2)
\sum_{j=1}^{14}(d_{x,j}^2 + d_{y,j}^2).
\label{eq:dispersion}
\end{equation}
The dispersion functions and beam positions were measured using 14 BPMs.
Additionally, the square value instead of the absolute value in
Eq.~(\ref{eq:dispersion}) was adopted to limit the outliers. The product of the
dispersion function and position was adopted to simultaneously reduce both the
dispersion function and beam-orbit residual. The dispersion function at each BPM
location was measured using the inevitably occurring beam-energy jitter.
Compared with the case where the dispersion function is measured by
intentionally changing the energy-adjustment knob, the method using energy
jitter enables measurements to be performed while the beam is being supplied to
the light source storage rings or SuperKEKB because the adjustment knob is
fixed. However, the energy-jitter method relies on randomly generated jitter and
requires a long measurement time to obtain sufficient resolution. In this
experiment, 100 data points were required after the pulsed magnet settings were
changed. Because the beam repetition rate was \SI{1}{\hertz}, a waiting period
of \SI{100}{\second} was permitted after the magnet settings were changed.

\subsection{Experimental results}\label{sec:result2}

\begin{figure}[hbt]
 \centering
 \includegraphics[width=\columnwidth]{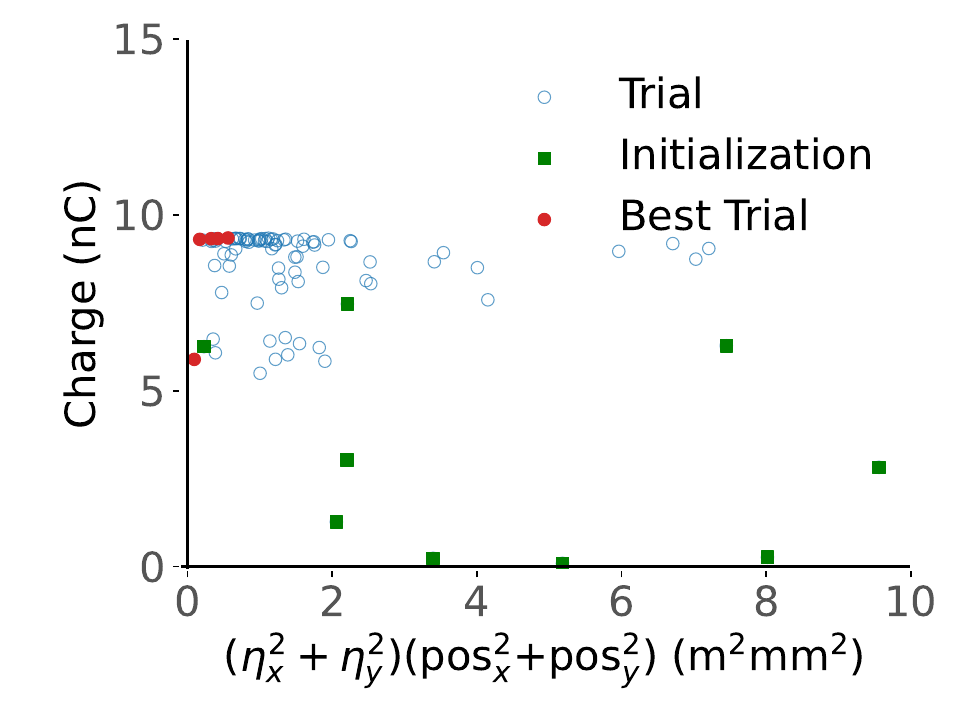}
 \caption{BoTorch result for multiobjective optimization on beam charge (vertical axis)
 and dispersion-position function (horizontal axis).}
 \label{fig:ParetoBo}
\end{figure}

Figure~\ref{fig:ParetoBo} shows the scatter plots of the obtained beam charge
vs. the dispersion-position function in Eq.~(\ref{eq:dispersion}) for each
trial. Measurements were performed on June 1, 2023, 2--4 am, with 100 trials
performed using the BoTorch algorithm.
The green squares indicate the 10 trials performed until the end of
initialization, the blue open circles indicate the 11th to 100th trials, and the
red dots are the Pareto-optimal solution set (five points in total). The
initialization explores the beam charge and dispersion-position function domain
more extensively than the pairs obtained after the initialization.
The beam charges were generally distributed above \SI{6}{\nano\coulomb}, and the
charge optimization was efficient. Meanwhile, the dispersion-position function
exhibited a tail exceeding \SI{2}{\square\meter\square\milli\meter}, which can
be further improved.
Although not used in this study, \textsc{optuna} implements a constrained
optimization function. If either the dispersion function or beam-orbit position
is constrained, then the optimization can focus of regions where the
dispersion-position function is small, e.g., smaller than
\SI{1}{\square\meter\square\milli\meter}.

\begin{figure}[hbt]
 \centering
 \includegraphics[width=\columnwidth]{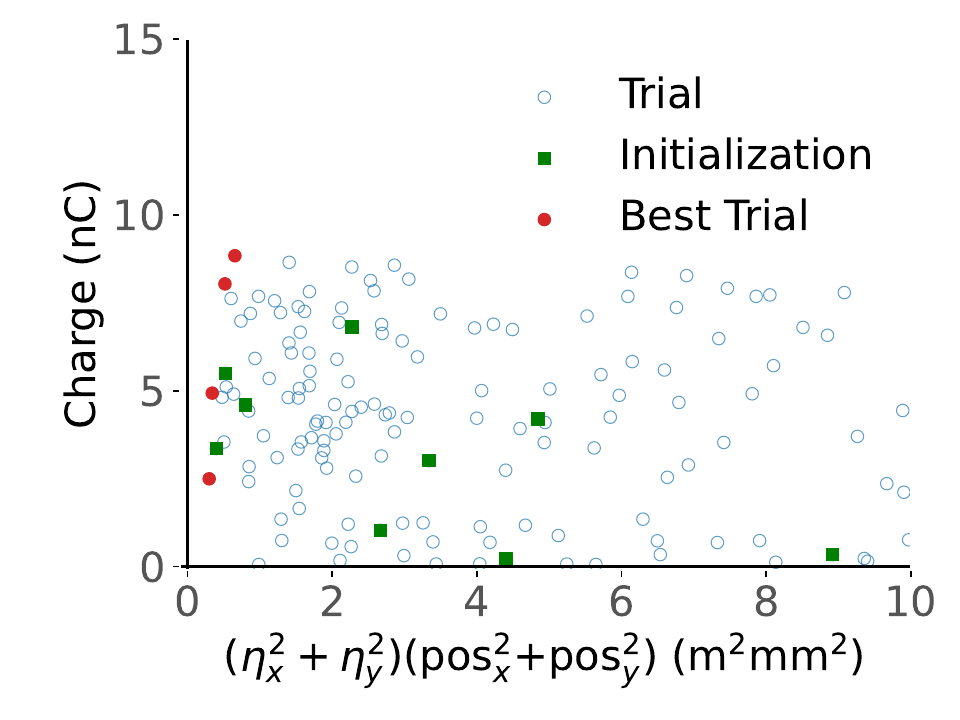}
 \caption{TPE result for multiobjective optimization on beam charge (vertical axis) and
 dispersion-position function (horizontal axis).}
 \label{fig:ParetoTPE}
\end{figure}

Figure~\ref{fig:ParetoTPE} shows a scatter plot of the beam charge vs.
dispersion-position function obtained using the TPE algorithm via 200 trials.
The result indicates an exploration-oriented optimization of both the beam
charge and dispersion-position function, unlike the BoTorch result presented in
Fig.~\ref{fig:ParetoBo}. This trend is consistent with TPE's focus on
exploration instead of exploitation, as discussed in Sec.~\ref{sec:result1}.
The TPE algorithm data were obtained on June 12, 2023, 6--10 pm; therefore, the
beam conditions may have changed since June 1 when we conducted the beam test
using the BoTorch algorithm.

As shown in Figs.~\ref{fig:ParetoBo} and \ref{fig:ParetoTPE}, we obtain
Pareto-optimal solutions for both algorithms. Only a few Pareto-optimal
solutions were available for 100 or 200 trials, thus clearly indicating a
low-cost performance in terms of the beamtime. The Pareto-optimal solution must
be obtained promptly for time-consuming measurements, e.g., more than 60 s per
measurement, as in the current tuning experiment.
In the future, we plan to test the efficiency of obtaining Pareto-optimal
solutions using a constrained optimization algorithm.

\section{Conclusions}\label{sec:conclusions}

We conducted beam-tuning experiments at the KEK Linac using Bayesian
optimization (BoTorch), a TPE, and the CMA-ES to determine the feasibility of
using machine learning, in particular optimization algorithms.

In a single-objective optimization experiment to maximize the electron-beam
charge from sector C to sector 1 of the Linac, the beam orbit was adjusted by
optimizing the applied current of six pulsed steering magnets.
The maximum beam charge obtained in the beam-tuning experiment was comparable to
that obtained in previous experiments. Approximately 35 trials (10 of which
pertained to initialization) were required to reach the maximum beam charge when
using the cold-starting BoTorch. On average, the TPE and CMA-ES achieved a lower
beam charge under cold starting compared with BoTorch, even after 100 trials.
Under warm starting, BoTorch and the CMA-ES showed excellent optimization
performance from the initialization phase, where the enqueued initial values
were utilized; however, the optimization performance of the TPE under warm
starting did not differ significantly from that of cold starting.
We conclude that the proposed optimization algorithm can replace manual tuning
by experts for beam-charge maximization using steering magnets.

Multiobjective optimization was tested to simultaneously maximize the beam
charge and minimize the dispersion-position function. The multiobjective
optimization task was to obtain as many effective Pareto-optimal solutions as
feasible. The dispersion function was measured using the inevitable beam energy
jitter. Two algorithms, i.e., BoTorch and the TPE, were used, and their results
were compared. As shown in the single-objective optimization results, the beam
charge vs. dispersion-position function distribution was exploitation-oriented
for BoTorch and exploitation-oriented for the TPE.
Both algorithms yielded four to five Pareto-optimal solutions over 100--200
trials. The efficiency of obtaining the optimal solution is essential for
applying multiobjective optimization to accelerator tuning, where each trial is
time consuming and hence expensive. In the next test, we shall introduce a
constrained optimization algorithm to improve the efficiency of obtaining the
optimal solution. In addition, the number of parameters shall be increased to
approximately 20 to assess the applicability of multivariable optimization to
beam tuning.

The machine-learning-based beam-tuning tool developed for this experiment will
be applied to the beam tuning of the SuperKEKB (e.g., beam injection from the
beam-transport line to the main ring, correction for horizontal and vertical
couplings, and adjustment of the beam collimator head position).

\begin{acknowledgements}
The authors would like to thank the members of the KEK Linac commissioning group
and the KEK Linac operators for their cooperation and assistance with our beam
study. Additionally, the authors thank the Belle II machine-detector interface
group for their cooperation and fruitful discussions.
\end{acknowledgements}

\bibliography{document}

\end{document}